\documentstyle[12pt,epsf,aaspp4]{article}

\def\spose#1{\hbox to 0pt{#1\hss}}

\def\etal{{\it et al. }}

\def\refindent{\par\noindent\parskip=4pt\hangindent=3pc\hangafter=1 }

\def\apj#1#2#3{\refindent#1,  {ApJ,\ }{#2}, #3}
\def\apjsup#1#2#3{\refindent#1,  {ApJS\ }{#2}, #3}

\def\apjlett#1#2#3{\refindent#1,  { ApJL,\  }{#2}, #3}

\def\mnras#1#2#3{\refindent#1,  { M.N.R.A.S., }{#2}, #3}

\def\aj#1#2#3{\refindent#1,  { AJ,\  }{#2}, #3}

\def\aa#1#2#3{\refindent#1,  { AA,\ }{#2}, #3}

\def\refbook#1{\refindent#1}

\def\ltsim{\mathrel{\spose{\lower 3pt\hbox{$\mathchar"218$}}
     \raise 2.0pt\hbox{$\mathchar"13C$}}}
\def\gtsim{\mathrel{\spose{\lower 3pt\hbox{$\mathchar"218$}}
     \raise 2.0pt\hbox{$\mathchar"13E$}}}
     
\def\apequal{\mathrel{\spose{\lower 1pt\hbox{$\mathchar"218$}}
     \raise 2.0pt\hbox{$\mathchar"218$}}}
\newbox\grsign \setbox\grsign=\hbox{$>$} \newdimen\grdimen \grdimen=\ht\grsign
\newbox\simlessbox \newbox\simgreatbox
\setbox\simgreatbox=\hbox{\raise.5ex\hbox{$>$}\llap
     {\lower.5ex\hbox{$\sim$}}}\ht1=\grdimen\dp1=0pt
\setbox\simlessbox=\hbox{\raise.5ex\hbox{$<$}\llap
     {\lower.5ex\hbox{$\sim$}}}\ht2=\grdimen\dp2=0pt

\def\etal{{\it et al.\ }}

\begin{document}
\input{psfig.sty}

\title{A Direct Detection of Dust in the Outer Disks of Nearby Galaxies}
\author{Amy E. Nelson}
\affil{Board of Astronomy and Astrophysics, Univ. of California, Santa Cruz, CA, 95064,
E-Mail: anelson@ucolick.org}
\author{Dennis Zaritsky}
\affil{UCO/Lick Observatory and Board of Astronomy and Astrophysics, Univ. of California, Santa Cruz, CA, 95064,
E-Mail: dennis@ucolick.org}
\author{and}
\author{Roc M. Cutri}
\affil{IPAC-Caltech, Mail Stop 100-22, Pasadena,
CA, 91125,
E-Mail: roc@ipac.caltech.edu}
\abstract{We measure the extent of 100$\mu$m galactic emission in two independent galaxy samples using the 
IRAS 100$\mu$m Sky Survey images and constrain the distribution of dust at large ($\ltsim$30~kpc) radii.  The first sample consists of 90
nearby (v~$<~$6000~km/s) galaxies from the RC3 catalog with similar
angular sizes and absolute luminosities ($5^{\prime}~\leq~$D$_{25}~\leq~10^{\prime}$ and
$-22.5~\leq~$M$_{B}~\leq~-18$) that are isolated in the 100$\mu$m
images.  The second sample consists of 24 local galaxies
(v~$<$~1500~km/s, $10^{\prime}~\leq~$D$_{25}~\leq~30^{\prime}$).  We rescale the
100$\mu$m images of these galaxies using their optical diameters,
D$_{25}$, rotate the images using their optical major axis position angle,
construct the mean and median image, and rebin the final
images into polar
coordinates to study the 100$\mu$m emission as a function of
radius and azimuthal angle.  We find that the
100$\mu$m emission extends at least to radii of 27~kpc (2$\sigma$ detection)
for the typical galaxy in the
$5^{\prime}~-~10^{\prime}$ sample and to 21~kpc (2$\sigma$ detection) in the
$10^{\prime}~-~30^{\prime}$ sample (H$_{0} = $75 km/s/Mpc).  In both samples,
the emission is
preferentially elongated along the optical major axis.  We fit an
exponential to the 100$\mu$m emission along the major axis and measure a scale
length of $2.5~\pm$~0.8~kpc (90\% confidence interval).  Using a
simple model that relates the far-IR emission to the stellar distribution, we examine the range of acceptable dust mass
distributions allowed by our data and conclude that the dust is more extended than the starlight.}


\section{Introduction}

Of the three principal baryonic components of any galaxy -- stars, gas,
and dust -- the distribution of dust is the least understood.  The
radial density
distribution of the stellar component in disk galaxies falls off
exponentially with a scale length ranging from $\sim$1~-~10~kpc
(de Jong 1996a,1996c) out to an
abrupt radial cutoff (van der Kruit and Searle
1981a, 1981b, 1982a, 1982b).  The radial density distribution of neutral hydrogen
is flat (e.g. Shaya and Federman 1987) out to a radial cutoff, while
that of
molecular gas is generally
exponentially decreasing (e.g. Young and Scoville 1991) even though
there is evidence for the presence of cold molecular gas at large
galactic radii (e.g. Lequeux \etal 1993; Fich \etal 1989; Wouterloot \etal 1990). 
These various distributions have been used to study the
formation of disks (cf. Fall \& Efstathiou 1980; Lin \& Pringle 1987), star formation in disks (Kennicutt 1989),
and the relation between the neutral hydrogen fraction and the ionizing intergalactic radiation field
(Maloney 1993).  Does dust
trace the stellar or gaseous components (either the atomic or molecular
phase)?  The resolution of this question may yield insight into
past star formation in galactic disks and into the energetics and transport
of the interstellar medium throughout disks.  The goal of this study is to
infer the average radial and angular distribution of dust at large radii
in galaxies, by measuring the 100$\mu$m emission from
ensembles of galaxies.

Existing studies of
dust in galaxies rely on one of two physical phenonemon:  (1)  the
extinction of light by dust grains (e.g. Zaritsky 1994; Block \etal 1994; Peletier \etal 1995; de Jong 1996b); and (2)  the emission of FIR/mm light by dust grains (e.g. Rice \etal 1988; Bothun \& Rogers 1992; Andreani \& Franceschini 1996;
Odenwald, Newmark \& Smoot 1996).  The application of either approach
becomes increasingly difficult at larger radii.  By measuring the
extinction of background galaxies observed through the halos of
two large, nearby spirals, Zaritsky (1994)
presented preliminary evidence of dust at radii of
$\sim$60~kpc.  Studies of the far-IR emission from galaxies have focused on dust
within the optically luminous disks of nearby galaxies of large
angular extent (Rice et
al. 1988; Bothun, Lonsdale, Rice 1989; Bothun \& Rogers 1992; Rice
1993).  Bothun and Rogers (1992) detected 100$\mu$m emission
out to approximately 30~kpc in 6 of the 28 galaxies in their sample.
However, the signal-to-noise is rather low in these outer regions, and there
is limited spatial information.  Reliable, general constraints on
the distribution of dust beyond $\sim$15~kpc are scarce.

We strive for sensitivity by coadding the 
100$\mu$m images
of 115 galaxies drawn
from the IRAS Sky Survey Atlas (ISSA; Wheelock \etal 1994) to measure the radial extent,
profile, and azimuthal distribution of the thermal dust emission.  We
choose to work with the longest wavelength IRAS images because we are interested in the
distribution of dust at large radii, which is most likely cold
($\ltsim 30K$) and
therefore has peak emission at $\lambda \gtsim 100\mu$m. 
In $\S$2, we discuss the selection 
criteria for the sample and control galaxies, and describe our
algorithm. In $\S$3, we present the results obtained from the two samples.
Finally, in $\S$4, we model the 100$\mu$m emission, and present
a simple model which constrains the dust distribution.

\section{The Data}

\subsection{Sample Selection}

\begin{figure}
\epsscale{1.0}
\plotfiddle{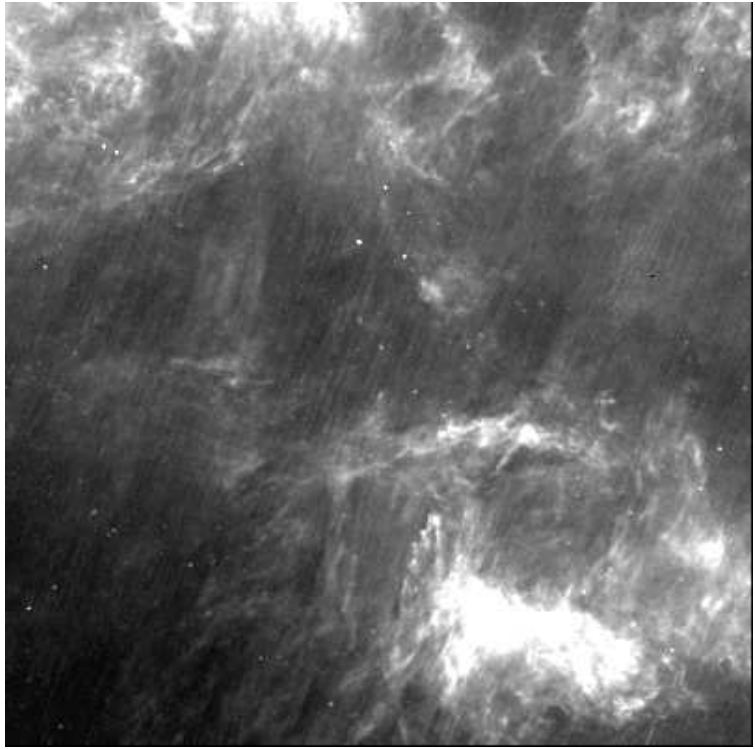}{4in}{0}{70}{70}{-210}{-90}
\caption{A $\sim 20^{\circ} \times 20^{\circ}$ section of the
ISSA Atlas near $(\alpha,~\delta) = (1^{h}20^{m},~0^{\circ})$ or
equivalently $({\it l},{\it b}) = (139.5^{\circ},~-61.6^{\circ})$.
The filamentary features of large angular extent are due to Galactic
cirrus, while the numerous small higher surface brightness knots are
predominantly galaxies.\label{f1}}
\end{figure}

Our measurement of the emission from cold dust in galaxies is
based on the combination of 100$\mu$m ISSA images (Wheelock \etal
1994) and basic galactic data from the Third Reference Catalog of Bright
Galaxies (RC3; de Vaucouleurs et
al. 1991).  We limit 
our study to Galactic latitudes $b \gtsim 15^\circ$ and $b \ltsim -40^\circ$ to minimize Galactic cirrus 
contamination and select candidates from the RC3 based solely on their coordinate
positions, regardless of other galactic properties such as morphology.
Figure~\ref{f1}~shows a 
$\sim20^\circ \times 20^\circ$ 
section of the 100$\mu$m Atlas near ($\alpha$,$\delta$) = (1$^h$20$^m$, 0$^\circ$) or equivalently
(${\it l},{\it b}$) = (139.5$^\circ$, $-$61.6$^\circ$).  The images have
1.5$^{\prime}$ pixels.  
The filamentary features of large angular extent 
are due to Galactic cirrus (Low \etal 1984).  The numerous small, high surface brightness knots are 
predominantly galaxies and are typically comparable in size to the point spread function 
of the survey (FWHM $\simeq~5^{\prime}$; ISSA Explanatory Supplement).  

The most serious difficulty faced by this study is the irregular structure of the Galactic foreground emission.
As illustrated in Figure~\ref{f1}, the contamination can be high even at high
galactic latitude.  To ensure that the final sample of galaxies do not
have close companions and that they lie in quiescent regions of the foreground
sky, we identify each candidate on the ISSA images using 
the RC3 coordinates and 
a gnomic coordinate projection centered on the central coordinates of our
35$^\circ$ survey images, and apply the following criteria:

\noindent  ${\it (1)~~Centering~Convergence:}$  The galaxies' 100$\mu$m luminosity 
centroids are calculated using the IRAF CENTER 
routine\footnote{IRAF is distributed by the 
National Optical Astronomical Observatories, which are operated
by AURA, Inc., under contract to the NSF.}. 
We reject galaxies with 100$\mu$m centroids that have internal
uncertainties $>$0.45$^{\prime}$ in either
x or y ($>$ 1/3 of a pixel), or that have 100$\mu$m centroids that differ by more than
2.25$^{\prime}$ ($\sim0.5\times$FWHM of a point source)
from the projected coordinates.  Typically, galaxies fail these
criteria if they have low signal-to-noise, a close companion, or lie in regions of
high cirrus contamination.  About 90$\%$ of the original 23660 possible target
galaxies are rejected due to this criteria.  

\noindent  ${\it (2)~~ Photometric~Convergence:}$  We perform aperture photometry on the remaining galaxy candidates using the 
IRAF PHOT routine and apertures with radii 
from 1.5$^{\prime}$ to 10.5$^{\prime}$ spaced by 1.5$^{\prime}$. We reject any galaxy whose magnitude does not converge 
on the grounds that the local background must be strongly affected by
cirrus emission.  This step excludes $\sim6\%$ of the remaining targets.

\noindent ${\it (3)~~Image~Edge~and~Bad~Pixel~Contamination:}$
Galaxies within 52.5$^{\prime}$ (a distance equal to the region of the
extracted subimage used in the analysis) of the image edge are
rejected.  In addition, the process of mosaicing smaller images to
produce the $\sim30^{\circ}~\times~30^{\circ}$ ISSA images can create
streaks of bad pixels at the interface of the sub-images.  Any
galaxies falling in regions of these bad pixel streaks are rejected.
Because both the shape of the final mosaic images and the initial
sub-images vary from image to image, this step is done interactively and excludes $\sim17\%$ of the remaining galaxies.

\noindent  ${\it (4)~~Multiple~Identifications:}$  Any 100$\mu$m
source that was associated with multiple optical sources was rejected;
excluding $\sim9\%$ of the remaining galaxies.

\noindent  ${\it (5)~~Close~Companion~Contamination:}$ 
To avoid contamination from 
close companion galaxies, we reject any galaxy with a cataloged 
companion within 30$^{\prime}$ 
projected separation.  This cut excludes
$\sim20\%$ of the remaining targets.  

\noindent  ${\it (6)~~Visual~Inspection:}$  To ensure 
that the final galaxy sample is not contaminated by either strong 
cirrus or close companions not listed in the RC3, we visually inspected 
each image.  Interactively, 
we only removed 3$\%$ of the remaining galaxies.  Overall, 95$\%$ of
the original galaxies are rejected due to the above cuts.   
 
This procedure yields a heterogeneous set of isolated galaxies, which span a
wide range of sizes, distances, luminosities, and inclinations.  For the
addition of galaxy images to be physically meaningful, we must
normalize their apparent angular scales and brightness using 
their isophotal diameters and their central surface brightnesses.  In
defining the final samples, we want to maximize the number of galaxies to
increase the sensitivity yet minimize the range of isophotal diameters and central surface
brightness to avoid large scaling factors that greatly magnify
the point spread function, PSF, (which could mask an otherwise
resolvable detection) and decrease the signal-to-noise ratio.
Therefore, we
impose two conditions.  First, the isophotal diameters of galaxies in
any particular sample cannot be scaled by more than a
factor of three.  We bin the samples into three groups according to $D_{25}$:  $2.5^{\prime} - 5^{\prime}$, $5^{\prime} - 10^{\prime}$, and
$10^{\prime} - 30^{\prime}$.  Second, the galaxies must be within a
limited range of intrinsic
luminosities.  We include those
galaxies in the $2.5^{\prime} - 5^{\prime}$ and $5^{\prime} - 10^{\prime}$ samples that
have $-18 > M_B > -23$ (H$_{0} = 75$~km/s/Mpc in all calculations).  The
galaxies in the $10^{\prime} - 30^{\prime}$ sample are sufficiently nearby that calculating luminosities using
recessional velocities, $V_{rad}$, is not accurate ($\langle V_{rad} \rangle =
748$~km/s), so no luminosity cuts are made.  

\begin{figure}
\epsscale{1.0}
\plotfiddle{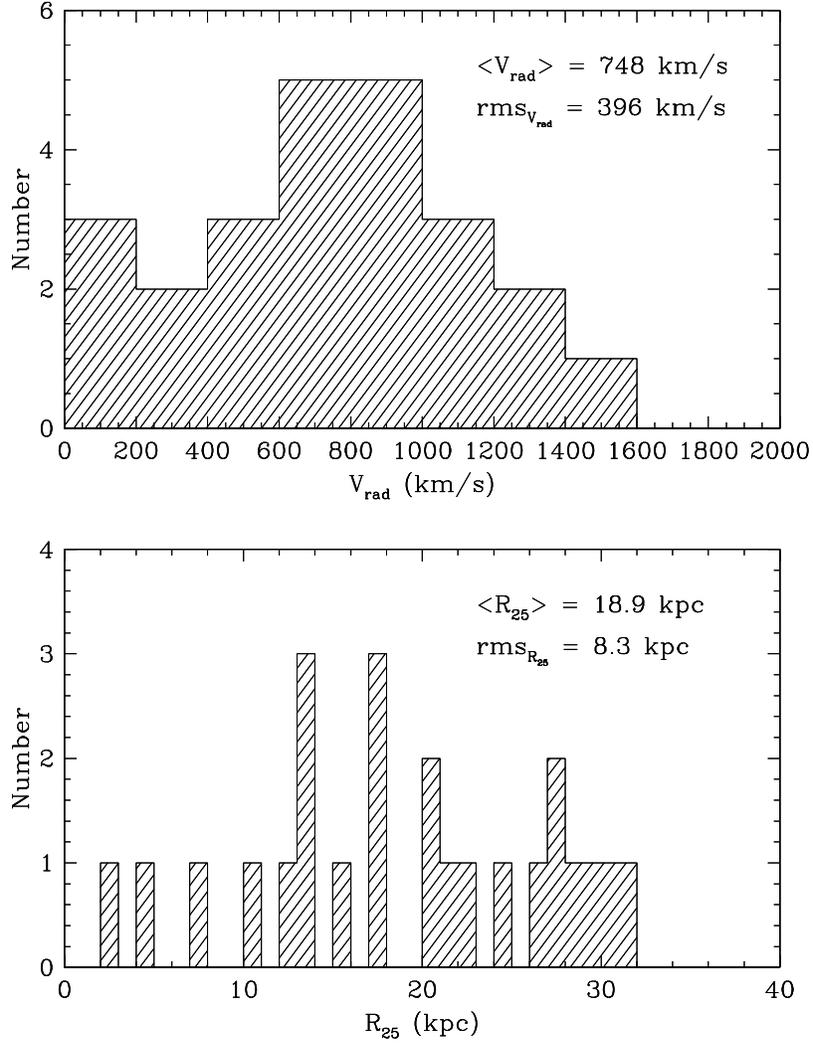}{4in}{0}{70}{70}{-210}{-70}
\caption{The distributions of recessional velocities and R$_{25}$
for the $10^{\prime} - 30^{\prime}$ sample of 24 galaxies.\label{f2}}
\end{figure}

\begin{figure}
\epsscale{1.0}
\plotfiddle{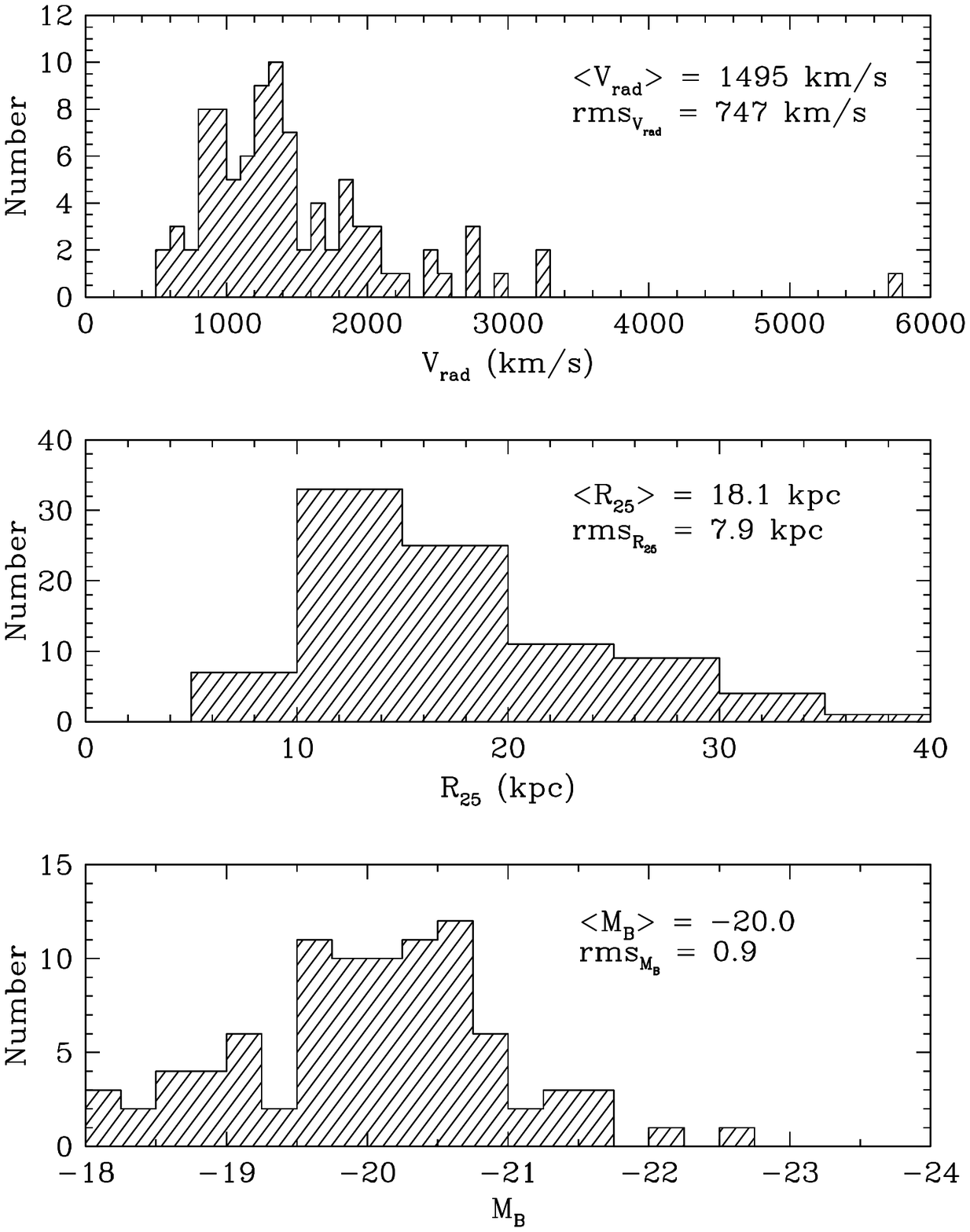}{4in}{0}{70}{70}{-210}{-70}
\caption{The distributions of recessional velocities, R$_{25}$,
and M$_{B}$ for the $5^{\prime} - 10^{\prime}$ sample of 91 galaxies.\label{f3}}
\end{figure}

\begin{figure}
\epsscale{1.0}
\plotfiddle{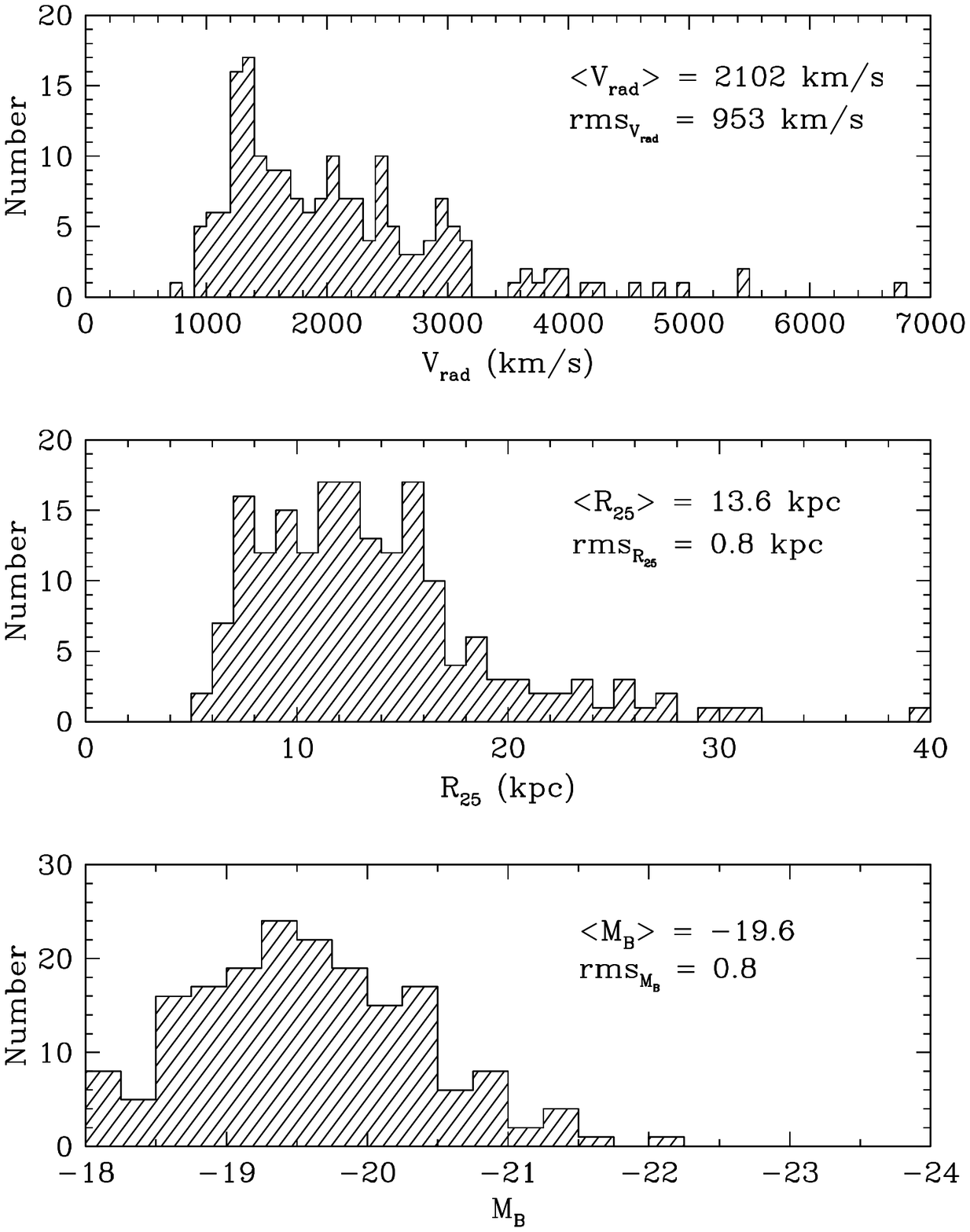}{4in}{0}{70}{70}{-210}{-75}
\caption{The distributions of recessional velocities, R$_{25}$,
and M$_{B}$ for the $2.5^{\prime} - 5^{\prime}$ sample of 184 galaxies.\label{f4}}
\end{figure}

\begin{figure}
\epsscale{1.0}
\plotfiddle{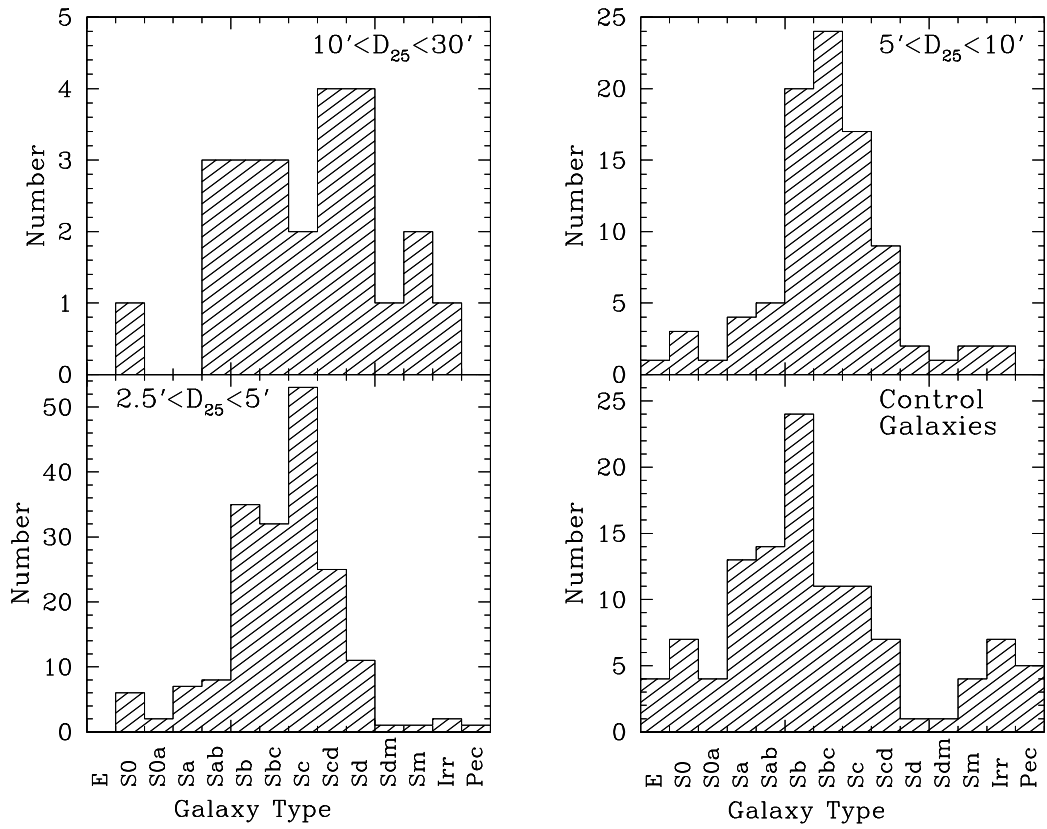}{4in}{-90}{70}{70}{-290}{390}
\caption{The distribution of galaxy type, as listed in the RC3,
for all three samples of galaxies and the control galaxies.\label{f5}}
\end{figure}

\begin{deluxetable}{lcccccc}
\footnotesize
\tablewidth{0pt}
\tablecaption{Sample Properties \label{t1}}
\tablehead{
Sample & \colhead{Number} & \colhead{D$_{25}$} & \colhead{V$_{rad}$ (km/s)} & \colhead{$\langle$V$_{rad} \rangle$
(km/s)} & \colhead{M$_{B}$ (mag)} & \colhead{$\langle$R$_{25}\rangle$
(kpc)}
}
\startdata
$10^{\prime} - 30^{\prime}$ & $24$ & $10^{\prime} - 30^{\prime}$ & $0 - 1600$ & $750$ &
\nodata & $18.9$\nl
$5^{\prime} - 10^{\prime}$ & $91$ & $5^{\prime} - 10^{\prime}$ & $500 - 3300$ & $1500$ &
$-18$ to $-23$ & $18.1$\nl
$2.5^{\prime} - 5^{\prime}$ & $184$ & $2.5^{\prime} - 5^{\prime}$ & $700 - 6800$ &
$2100$ & $-18$ to $-23$ & $13.6$\nl
Control Galaxies & $200$ & $0^{\prime} - 1.5^{\prime}$ & $1000 - 16000$ & $5800$ & $-18$
to $-23$ & \nodata\nl
Control Stars & $30$ & \nodata & \nodata & \nodata & \nodata & \nodata\nl
\enddata
\end{deluxetable}

The final samples
consist of 24 galaxies in the $10^{\prime} - 30^{\prime}$ sample, 91
galaxies in the $5^{\prime} - 10^{\prime}$ sample, and 184 galaxies in the $2.5^{\prime} - 5^{\prime}$ sample.  The distributions of recessional velocities,
R$_{25}$, and M$_{B}$ for each sample, if available, are shown in
Figures~\ref{f2},~\ref{f3},~\ref{f4}.  A
summary of the mean properties of each sample is given in Table~\ref{t1}.
Although no selection criteria were imposed regarding galaxy type, the
samples are predominantly galaxies classified as
type Sd to Sa in the RC3, as shown in Figure~\ref{f5}.  This is an
indirect result of selecting galaxies from the 100$\mu$m images, a
spectral regime where emission from dusty late-type galaxies dominates
over the relatively dust-free early-type galaxies.  Notice that the
galaxies in the $10^{\prime}~-~30^{\prime}$ and $5^{\prime}~-~10^{\prime}$ samples
appear to contain nearly identical galaxies because they both have
$\langle R_{25} \rangle \sim 18.5$~kpc, while the $2.5^{\prime}~-~5^{\prime}$ sample
appears to be dominated by galaxies that are physically smaller,
$\langle R_{25} \rangle \sim 13.5$~kpc.
Finally, as we discuss below, we take care to select a control sample with the same
quantitative criteria as used to select the original samples.

\subsection{Coadded Images and the Point Spread Function}

The exact procedure we use to prepare the galaxy images to be coadded consists of
the following steps: 

\noindent (1) we extract a subimage of $121.5^{\prime} \times 
121.5^{\prime}$ (81 pixels)
centered on each galaxy using the 
100$\mu$m galaxy centroid;

\noindent (2) this subimage
is rotated and scaled using IRAF's IMLINTRAN routine and the optical
position angle and D$_{25}$ from the RC3, while
conserving total flux;

\noindent (3) we subtract
a mean sky value measured in an annulus defined by $30^{\prime} < R < 52.5^{\prime}$
, where R is the radius from the target galaxy; and

\noindent (4) we normalize the flux, as measured by our aperture photometry, so that the galaxy's central
surface brightness is equal to unity. 

\noindent Upon completion,
the resulting images of each galaxy are combined into a mean and median image,
applying $10\sigma$ clipping to remove spurious pixels.
The outer border (width = 10 pixels) of the resulting co-added image is fit by a plane
to remove any large scale sky variations.  The image is then rebinned into polar
coordinates, which allows a straightforward comparision of the radial extent
of the 100$\mu$m emission as a function of polar angle (e.g. Fig 7).  Finally, the sky is resubtracted (always a small subtraction, $<
1\%$) and the central surface brightness is renormalized (always a
small division factor, resulting in $< 1\%$ effect). The
result of this procedure is an image in polar coordinates that
is the sum of 184 galaxies in the $2.5^{\prime} - 5^{\prime}$ sample, 91 galaxies in
the $5^{\prime} - 10^{\prime}$ sample, and 24 galaxies in the $10^{\prime} - 30^{\prime}$ sample from
which we measure the shape and extent of
the 100$\mu$m emission.

We do not attempt any correction for galaxy inclination angle.
For all but the most nearby, face-on galaxies, the 5$^{\prime}$ PSF of
the ISSA images equals or exceeds the galaxies' minor axes.  However,
the major difficulty with any attempt at deprojection is the complex
background component that cannot be removed on a galaxy-by-galaxy
basis.  The other option to address this issue, exclusion of lower
inclination galaxies, is also not tenable.  This option substantially
decreases the sample size and our ability to detect emission at large
radii.  Therefore, we have chosen to apply no inclination correction,
but only present quantitative results from data taken along the major
axis of the final summed image, which is least affected by inclination.  

Most, if not all, of the 100$\mu$m emission from galaxies in the
samples are dominated by emission from the unresolved or marginally
resolved inner portions of the disks.  Consequently, our ability to
detect extended 100$\mu$m emission is highly dependent on the
determination of an accurate 100$\mu$m point spread function.  At a
distance of 25
Mpc, $\langle$D$_{25} \rangle$
corresponds to 5$^{\prime}$.  The IRAS 100$\mu$m PSF is 
$\sim$5$^{\prime}$ on the longest side, but is asymmetric and varies 
spatially across the images because of survey coverages and thus, is too complicated to model. We
will rely on an empirical determination from two independent sets of
unresolved sources.

\begin{figure}
\epsscale{1.0}
\plotfiddle{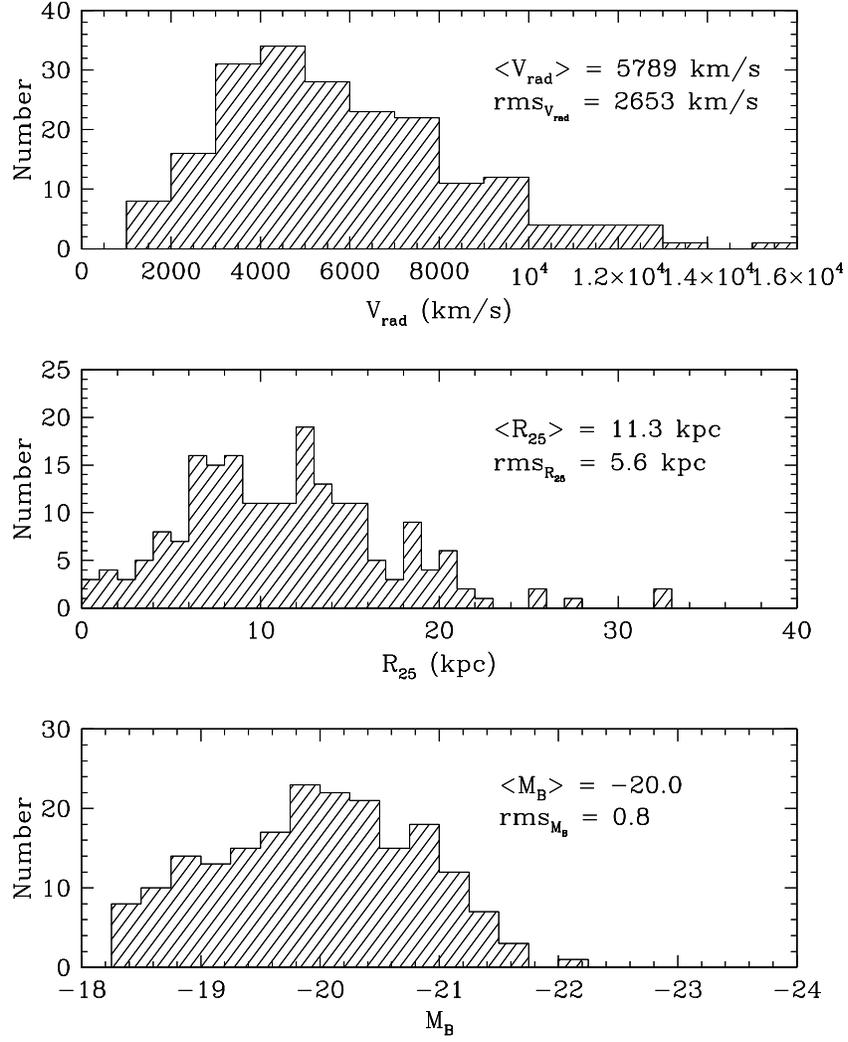}{4in}{0}{70}{70}{-210}{-70}
\caption{The distributions of recessional velocities, R$_{25}$,
and M$_{B}$ for the control sample of 200 galaxies.\label{f6}}
\end{figure}

To determine the point spread function in an empirical manner that is 
entirely consistent with the collection and analysis of our sample, we select two independent sets of unresolved
sources: stars detected in all four bands in the IRAS Faint Source Catalog; and compact
galaxies ($D_{25} \leq 1.5$$^{\prime}$) from the RC3 catalog.  As long
as stars do not have extended structure (such as debris disks or
envelopes) then they are
truly point sources, and so, apparently an ideal control set to
determine the PSF empirically.  Such extended structure is
characterized by an IR excess, and therefore we select stars with
F$_{12}~>~$F$_{60}$, where F$_{12}$ and F$_{60}$ are the fluxes at
12$\mu$m and 60$\mu$m, respectively.  Unfortunately, there are too few stars (30 that survived our selection
criteria) to compare to the more numerous $2.5^{\prime}~-~5^{\prime}$ and $5^{\prime}~-~10^{\prime}$
samples, and the stars 
have a much higher central surface brightness than the galaxies, which
may be an issue with detector hysteresis.  However, because the
$10^{\prime}~-~30^{\prime}$ sample is few in number (N$_{10^{\prime}-30^{\prime}}$
= 24) and those galaxies in the sample are bright, we use the stars
to determine the PSF to use for comparison to this sample.  The
compact galaxies are generally
distant ($\langle$V$_{rad}\rangle$ = 5789 km/s) and unresolved ($D_{25} \leq 1.5^{\prime}$).
These control galaxies are much more numerous ($N_{con} =
200$) than the stars, and have brightness characteristics that are
very well matched to the two sets of sample galaxies, $2.5^{^{\prime}}~-~5^{^{\prime}}$ and
$5^{\prime}~-~10^{\prime}$, to which they are compared.  Figure~\ref{f6} shows the
distribution of recessional velocities, $R_{25}$, and $M_B$ for the
control galaxies.  As is the case for the sample galaxies, the control galaxies
are predominantly of types Sc and earlier (see Figure~\ref{f5}).  This set of control
objects results in a conservative estimate of the 
PSF because the dust emission from these galaxies is not necessarily unresolved.  See Table~\ref{t1}
for a summary of the mean properties of the two control samples.

We calculate 
the PSF using these two control samples and Monte Carlo techniques
to reproduce the range of spatial rescalings and rotations applied to the data.
Our procedure includes: 

\noindent (1)  the extraction of a $121.5^{\prime} \times 121.5^{\prime}$
control galaxy sub-image, randomly selected from the complete control
set;

\noindent (2)  the control sub-image is spatially rescaled and rotated
using parameters drawn from a corresponding sample galaxy;

\noindent (3)  the control galaxy is sky-subtracted and
normalized so that its central
surface brightness equals unity, as is done for the sample galaxies.

\noindent These steps are repeated until the number of randomly selected control
galaxies equals the number of sample galaxies.   
The control objects are then 
coadded using the same algorithm as outlined previously for the sample
galaxies to produce an image of the PSF.  For the $2.5^{\prime} - 5^{\prime}$ and $5^{\prime} - 10^{\prime}$
samples, we repeat this procedure 20
times to get a mean representation of the point spread function and a
measurement of the 
uncertainty in the mean PSF.  For the $10^{\prime} - 30^{\prime}$ sample, the number of control stars is
comparable to the number of sample galaxies and thus we only repeat
the above Monte Carlo procedure 10 times to derive a corresponding mean PSF.

\section{Results}

\begin{figure}
\epsscale{1.0}
\plotfiddle{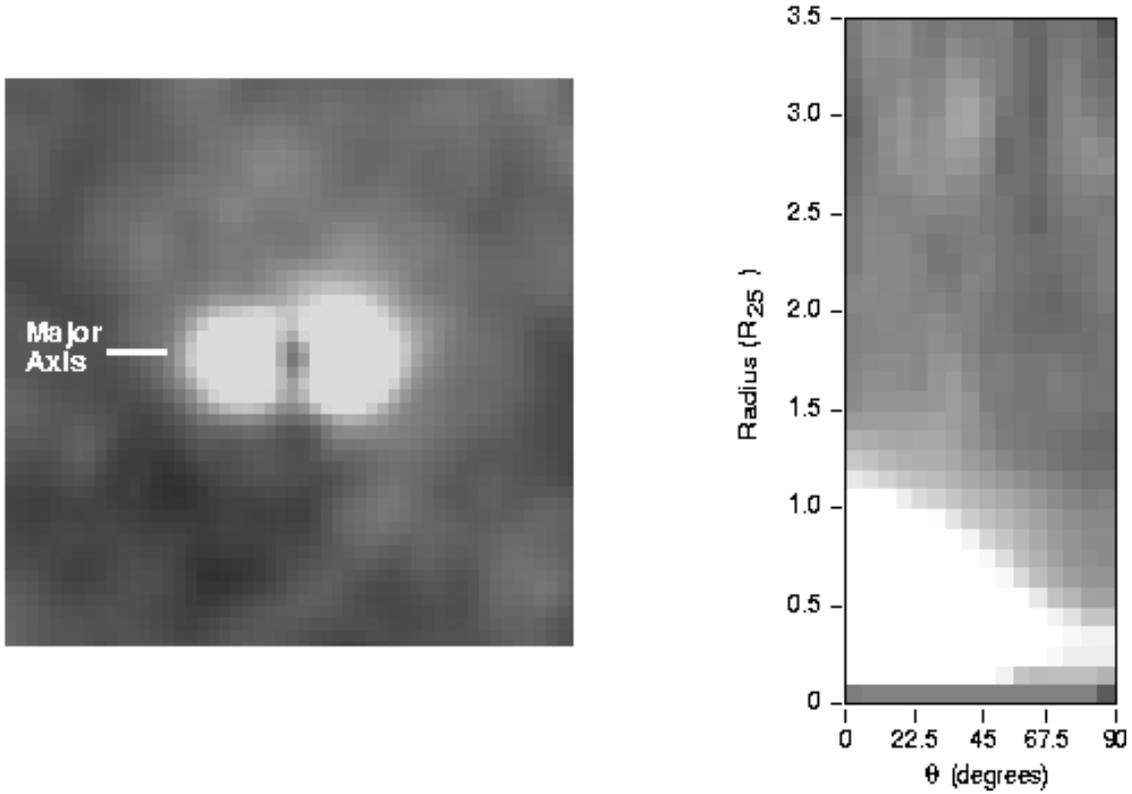}{4in}{-90}{85}{85}{-400}{465}
\caption{Residual images for the $10^{\prime} -
30^{\prime}$ sample, before $({\it left~panel})$ and after $({\it
right~panel})$ rebinning into polar coordinates.  The left panel image is $\sim 45^{\prime} \times 45^{\prime}$.
The residual is determined by calculating the mean image of the 10
Monte Carlo realizations of the control PSF and subtracting the mean
PSF from the coadded sample of galaxies image.  Lighter shades
correspond to higher flux.  100$\mu$m emission is visible to a radius
of $\sim$1.1R$_{25} \simeq 21$~kpc and is most extended along the major
axis of the disk, demonstrating that 100$\mu$m emitting dust is flattened
into a disk.\label{f7}}
\end{figure}

Any 100$\mu$m flux in excess of that expected from a point source is a
direct detection of extended emission from galaxies.  To determine if
excess emission exists, we construct residual images by subtracting
the PSF image from the coadded sample image.  The residual image for
the $10^{\prime} - 30^{\prime}$ sample is presented in Figure~\ref{f7}, where the
left hand and right hand panels are the residual images before and
after rebinning into polar coordinates, respectively.  The residual image was produced by combining the sample galaxy
images using a median filter.  The right hand panel shows
100$\mu$m emission that is easily visible to a
radius of $\sim$~1.1~R$_{25}~\simeq$~21~kpc.  Also, both panels illustrate that the residual emission is most extended along
the optical major axis of the
disk, thus indicating that the dust responsible for the 100$\mu$m
emission is flattened into a disk.

\begin{deluxetable}{lcc}
\tablewidth{0pt}
\tablecaption{100$\mu$m Emission Detections \label{t2}}
\tablehead{
\colhead{Sample} & \multicolumn{2}{c}{100$\mu$m Emission
Detection}\nl
\nl
\colhead{} & \colhead{2$\sigma$ Detection} & \colhead{1$\sigma$ Detection}
}
\tablecolumns{3} 
\startdata
10$^{\prime}$ - 30$^{\prime}$ & 1.1R$_{25}$ & 1.2R$_{25}$\nl
& 21~kpc & 23~kpc\nl
\nl
5$^{\prime}$ - 10$^{\prime}$ & 1.5R$_{25}$ & 1.8R$_{25}$\nl
& 27~kpc & 33~kpc\nl
\nl
$2.5^{\prime} - 5^{\prime}$ & \multicolumn{2}{c}{no emission detected}\nl
\enddata
\end{deluxetable}

\begin{figure}
\epsscale{1.0}
\plotfiddle{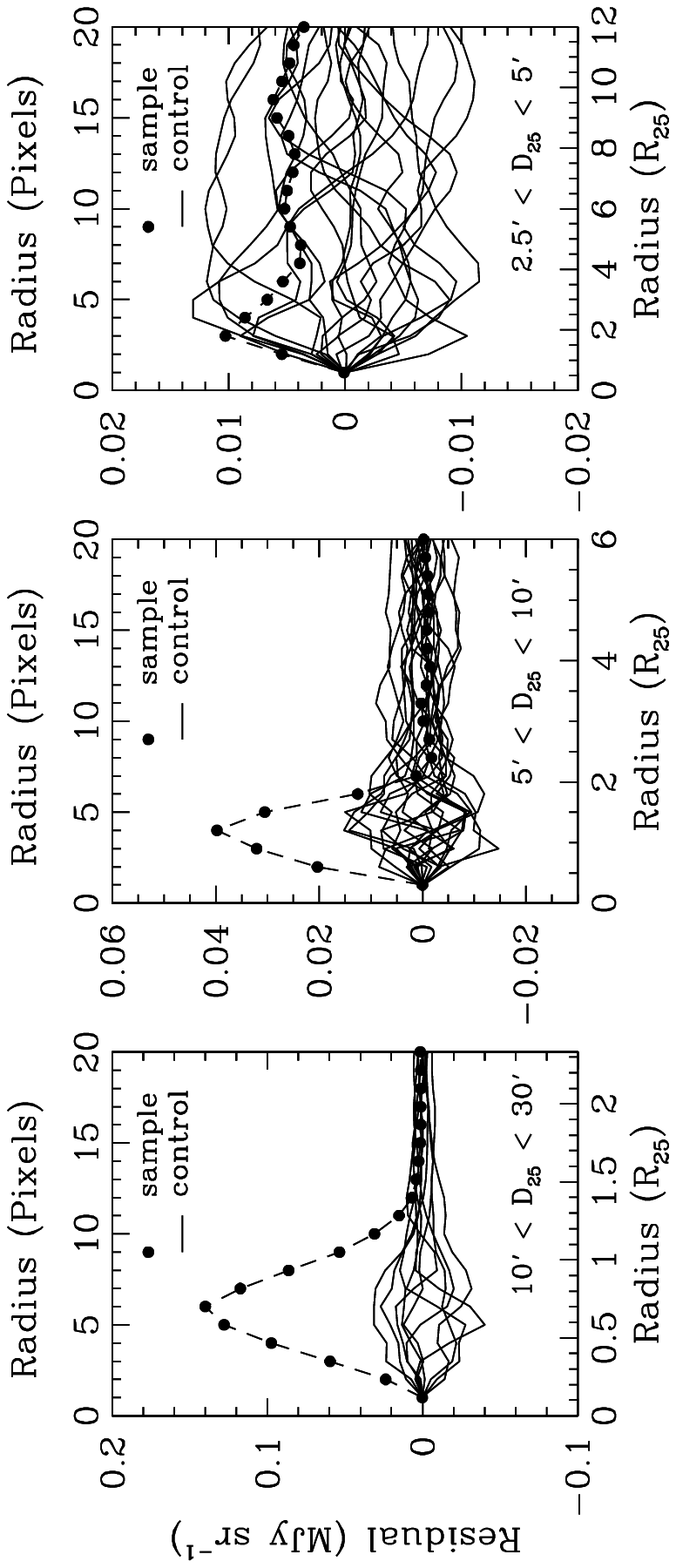}{4in}{-90}{85}{85}{-345}{360}
\caption{The surface brightness profile (averaged over
$0^{\circ} \leq \theta \leq 22.5^{\circ}$, where $\theta$ is the azimuthal angle such that 0$^{\circ}$
is along the disk major axis) for the $10^{\prime} - 30^{\prime}$ sample $({\it
left~panel})$, the $5^{\prime} - 10^{\prime}$ sample $({\it middle~panel})$, and the
$2.5^{\prime} - 5^{\prime}$ sample $({\it right~panel})$.  The mean PSF has been subtracted from the coadded sample galaxies and also
from each individual Monte Carlo realization of the control.  The
scatter of the various realizations of the control residuals are indicative of
the uncertainties and any emission above this scatter is a significant detection.  Extended 100$\mu$m emission is seen to radii of
1.1R$_{25} \sim 21$~kpc for the $10^{\prime} - 30^{\prime}$ sample and 1.5R$_{25} \sim
27$~kpc for the $5^{\prime} - 10^{\prime}$ sample at the $2\sigma$ level.\label{f8}}
\end{figure}

The radial distribution of the residual 100$\mu$m emission (averaged over $0^{\circ} \leq |\theta| \leq
22.5^{\circ}$, where $\theta$ is the angle such that
0$^{\circ}$ is along the disk major axis) for all three samples is shown in Figure~\ref{f8}.  The
mean PSF has been subtracted from the coadded
sample galaxies (combined using a median filter) and also, for comparision, from each individual Monte Carlo realization
of the control.  Uncertainties in the residual surface brightness
profile are calculated from the rms scatter of the various Monte Carlo realizations of the control
samples.  At the
$2\sigma$ level, 100$\mu$m emission extends to a radius of 1.1R$_{25}
\simeq$ 21~kpc for the $10^{\prime}~-~30^{\prime}$ sample and
to 1.2R$_{25} \simeq$ 23~kpc at the $1\sigma$ level (Fig~\ref{f8}, left panel).  For the $5^{\prime} - 10^{\prime}$ sample, residual 100$\mu$m
extended emission is detected to a radius of 1.5R$_{25} \simeq$ 27~kpc at the
2$\sigma$ level and to 1.8R$_{25} \simeq$ 33~kpc at the $1\sigma$
level (Fig~\ref{f8}, middle panel).  The $2.5^{\prime} - 5^{\prime}$ sample
(Fig~\ref{f8}, right panel) shows no significant extended emission at any
radii.  This null detection is consistent with our two previous
detections because the apparent angular
size of the 100$\mu$m detection in the two nearby samples is not resolvable at the mean distance
of the $2.5^{\prime} - 5^{\prime}$ sample and thus
serves as a check on our method, demonstrating that a spurious
detection is not produced by the method.  These results are summarized
in Table~\ref{t2}.  Combination using an average filter produces
virtually identical results (at the 2$\sigma$ level:  1.1R$_{25}
\simeq$ 21~kpc for the $10^{\prime} - 30^{\prime}$ sample and 1.8R$_{25} \simeq$
33~kpc for the $5^{\prime} - 10^{\prime}$ sample) and therefore are not shown explicitly.
Because the results are the same using either the average or median
combination, we are assured that these
detections are not due to one (or a few) dominant galaxies.  Rather,
extended 100$\mu$m emission is a typical feature of disk galaxies.

\section{Discussion}

We now investigate the
distribution of the 100$\mu$m emission in more detail and attempt to
constrain the distribution of dust.  First, we
examine the radial shape of the far-IR surface brightness profile and
compare it to that of the underlying stellar component.  Second, we use a simple model to convert the surface brightness
profile to a dust mass profile.  Finally, we briefly
discuss the implications of our derived dust mass distribution.

\subsection{100$\mu$m Surface Brightness Profile}

We have detected 100$\mu$m emission at radii of 21 kpc
for the $10^{\prime} - 30^{\prime}$ sample and 27 kpc for the $5^{\prime} -
10^{\prime}$ sample at the 2$\sigma$ level, but
what is the shape of that surface brightness profile?
Although the resolution of our data is coarse (HWHM$_{10^{\prime}-30^{\prime}}$
$\simeq$ 9.5 kpc and HWHM$_{5^{\prime}-10^{\prime}}$ $\simeq$ 19 kpc), the pixel scale
(1.9 kpc pixel$^{-1}$ for the $10^{\prime} - 30^{\prime}$ sample and 5.7 kpc pixel$^{-1}$
for the $5^{\prime} - 10^{\prime}$ sample) is sufficiently small to provide
for several pixels of coverage over the detection.  We do not resolve extended
emission along the disk minor axis and therefore, we confine ourselves
to parametrizing the major axis profile.  However,  we can conclude
that the 100$\mu$m emission is confined to within 9.5 kpc of the disk
plane for
the $10^{\prime} - 30^{\prime}$ sample and within 19 kpc for the
$5^{\prime} - 10^{\prime}$ sample (independent of inclination because
these are upper limits). 

\begin{figure}
\epsscale{1.0}
\plotfiddle{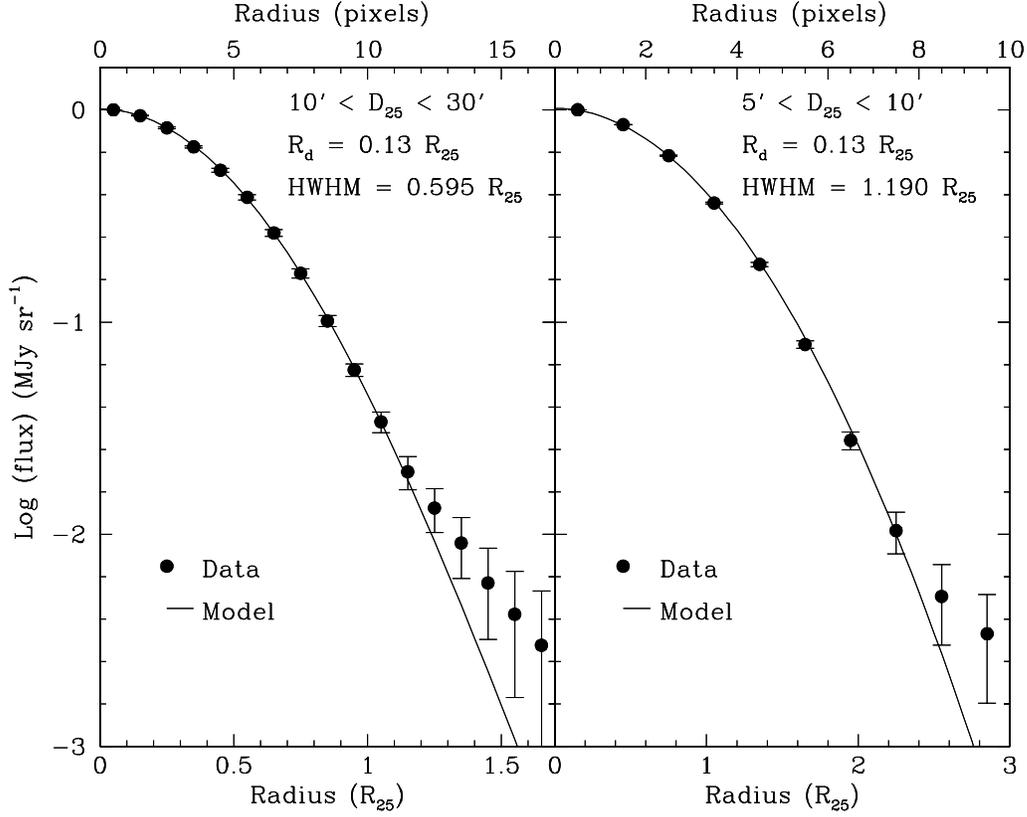}{4in}{-90}{70}{70}{-270}{375}
\caption{The simultaneous best fit to the 100$\mu$m surface brightness
profiles of the $10^{\prime} - 30^{\prime}$ sample $({\it left~panel})$ and the $5^{\prime} -
10^{\prime}$ sample $({\it right~panel})$, before renormalizing the errors.  The
surface brightness profiles are modelled by an exponential convolved
with a Gaussian (see text for details).  The $1\sigma$ error bars plotted are
derived from the rms scatter of the mean of the various Monte Carlo
realizations of the control samples.  \label{f9}}
\end{figure}

Along the direction of maximum emission ($0^{\circ} \leq
|\theta| \leq 22.5^{\circ}$), we fit an exponential radial profile to
the surface brightness, a model for the underlying light distribution, convolved with a
Gaussian, a model for the PSF.  Because we do not know the
effective PSF or the exponential scale length of the 100$\mu$m
emission, both the HWHM of
the Gaussian and the exponential scale length, r$_{d}$ are free
parameters.  We
simultaneously fit the data from both samples to the convolved
function by minimizing $\chi^{2}$ (and imposing the conditions that HWHM$_{5^{\prime}-10^{\prime}}$
= 2.0 $\times$ HWHM$_{10^{\prime}-30^{\prime}}$ and that the exponential scalelengths be
the same).  Figure~\ref{f9} demonstrates our best fit,
which is obtained for HWHM$_{10^{\prime}-30^{\prime}}$ = 0.595R$_{25}$ = 11.0~kpc
and $r_{d}$ = 0.13R$_{25}$ = 2.4~kpc, with a reduced $\chi^{2}$ = 2.84.  For
the majority of the 100$\mu$m surface brightness profile, the best fit is generally within
the 1$\sigma$
error bars, which are derived from the rms scatter of the mean
of the various Monte Carlo realizations of the control samples.  More
significant deviations occur in the outer regions of the profile, when
the flux values approach sky levels ($\sim$
1R$_{25}$ for the $10^{\prime} - 30^{\prime}$ sample and $\sim$ 2R$_{25}$ for
the $5^{\prime} - 10^{\prime}$ sample).  Although the errors are
large, we may be seeing evidence for a second, more extended component
at large radius (r$\gtsim$25~kpc).

The task of determining the uncertainties of the parameters is more
difficult than that of identifying the best parameters.  Although some parameters yield good fits that cannot be rejected at a
confidence level $>$90$\%$, neither model has a reduced $\chi^{2} \sim 1$.
Either the model is not entirely representative and there is a
nonexponential tail to the dust emission, or the uncertainties are
underestimates.  We conservatively suggest that $\chi^{2}$ is inflated.  As a result
of details of the polar rebinning procedure, pixels at small radii are
not independent (i.e. for small $r$, one $(x,y)$ pixel
contributes to different $(r,\theta)$ pixels after
rebinning).  Therefore, the calculated rms scatter for pixels at small
$r$ does not fully reflect the uncertainties.  To compensate for this
effect, we now renormalize
the errors in the following manner.  Because the underestimation of uncertainties
is more severe at small radii than at large, we renormalize the
errors using a factor $\propto$ A/r and define A to result in
$\chi^{2} = 1.0$ while fixing the model Gaussian to have 0.585R$_{25} \leq$
HWHM$_{10^{\prime}-30^{\prime}} \leq$ 0.595R$_{25}$, drawn from the
best fit (the IPAC estimated PSF
has HWHM$_{10^{\prime}-30^{\prime}} \sim$ 0.50R$_{25}$).  Using the
renormalized errors and fixing the range of the HWHM, we again
simultaneously fit the models to the data from both samples and find
the following parameter range acceptable within the 90$\%$
confidence level:  0.09R$_{25} \leq$ r$_{d} \leq$ 0.18R$_{25}$ or 1.7~kpc
$\leq$ r$_{d} \leq$ 3.3~kpc.

\begin{deluxetable}{lcc}
\tablewidth{0pt}
\tablecaption{Best Fit\tablenotemark{a}~~Parameters for 100$\mu$m
Surface Brightness Profile Fitting \label{t3}}
\tablehead{
Sample & \colhead{HWHM$_{10^{\prime}-30^{\prime}}$} & \colhead{r$_{d}$}
}
\startdata
10$^{\prime}$ - 30$^{\prime}$ & 0.520R$_{25}$ - 0.650R$_{25}$ &
0.05R$_{25}$ - 0.20R$_{25}$\nl
& 9.6~kpc - 12.0~kpc & 0.9~kpc - 3.7~kpc\nl
\nl
5$^{\prime}$ - 10$^{\prime}$ & 0.565R$_{25}$ - 0.610R$_{25}$ &
0.05R$_{25}$ - 0.20R$_{25}$\nl
& 10.5~kpc - 11.3~kpc & 0.9~kpc - 3.7~kpc\nl
\nl
Simultaneous fits to & 0.585R$_{25}$ - 0.595R$_{25}$ & 0.09R$_{25}$ -
0.18R$_{25}$\nl
both samples & 10.8~kpc - 11.0~kpc & 1.7~kpc - 3.3~kpc\nl
\enddata
\tablenotetext{a}{Within 90\% confidence level}
\end{deluxetable}

One important result that may be casually overlooked is the significance
of simultaneously fitting the data from both samples to the same
exponential.  One might naively expect that dust should be detected in
either sample out to the same radius, particularly since, as pointed
out previously, the galaxies
in these two samples do appear to be the same type of galaxies
($\langle$R$_{25}\rangle \sim 18.5$~kpc for both samples).  However, due to the differences
in sample size, the mean distance of galaxies in the two samples, and
the effect of the PSF on samples of different angular size,
the sensitivities are not the same.  Separate fits to the two samples
yield the $90\%$ confidence parameter ranges given in Table~\ref{t3}.  The range of
acceptable parameter space allowed by fitting the two samples separately
is consistent with that allowed by simultaneously fitting both samples
and with each other. 

How does our derived 100$\mu$m emission scalelength compare to that
of the stellar component in a typical disk galaxy?  The most recent
and complete study of stellar disks is that of de Jong (1996a, 1996c), who uses
near-infrared and optical broadband surface photometry of 86 face-on,
disk dominated galaxies to investigate the distribution of disk and
bulge parameters.  Because K-band light is less affected by dust
obscuration than B-band light, we choose to compare our result to the
distribution of K scalelengths which range from $\sim$~1~-~10~kpc,
with a median of $r^{K}_{d}=$3.8~kpc~,
and a rms dispersion of 2.4~kpc (for H$_{0} = 75$~km/s/Mpc).  However, the distribution of
scalelengths is a positive definite quantity with a significant
high-end tail that is not described
by a Gaussian.  Therefore, the median is biased to a larger value
than the peak of the distribution, which is quite broad and is in the range
$\sim$ 2~-~4~kpc ($40\%$ of the galaxies have scalelengths in this
range).  This range overlaps the $2.5\pm0.8$~kpc range of 100$\mu$m scale
lengths that we measure.  Because de
Jong finds no correlation of scalelength with galaxy type, any differences in the distribution of galaxy type between the optical
study and our samples is unimportant.  We conclude that the
derived 100$\mu$m scalelength is most likely not much larger than that of the
underlying old stellar component and is consistent with being equivalent
to or up to a factor of two smaller than that of the stellar distribution. 

\subsection{A Simple Model for the Dust Distribution}

The characteristic scale length of the 100$\mu$m emission
derived in the previous section can be used, in conjunction with
simple modelling, to place a limit on the global dust
mass distribution.  Because we have data in a single passband with
poor spatial resolution, we follow the procedure
of Bothun $\&$ Rogers (1992), who use a one-component model (a single
blackbody with characteristic temperature) to
describe the far-IR emission from dust, based on 60$\mu$m and 100$\mu$m
IRAS images.  The one-component assumption
is questionable because of different
temperature dust components needed to
fit observed far-IR spectral energy distributions.  Some find that
multi-component models are needed to explain their far-IR observations
(e.g. Helou 1986; Rowan-Robinson \& Crawford 1989; Andreani \& Franceschini 1996), while
others concur with Bothun \& Rogers (1992) and find that their far-IR spectra can be well fit by a single component
model (e.g.  Sodroski \etal 1994; Clements, Andreani and Chase 1993;
Odenwald, Newmark and Smoot 1996).  Using DIRBE all-sky data, Sodroski \etal (1994) showed that a
ubiquitous far-IR component can be detected as a single component at $60 \mu m < \lambda <
240 \mu m$ with $17 K < T_{d} < 22 K$ that is closely
associated with the far-IR cirrus discovered by IRAS at 100 $\mu$m.
Studies of IR selected galaxies find that the galaxies' continua are
consistent with a single component model with
25~$\ltsim~$T$_{d}~\ltsim$~35, and a peak emission near
100$\mu$m (Clements, Andreani and Chase 1993; Odenwald, Newmark and Smoot
1996).  While dust certainly exists at a variety of temperatures, we
use a simple one component model because of the limited temperature
information provided by our single band data, and the evidence that
the far-IR continua of the 100$\mu$m emitting dust is well fit by a
single temperature blackbody in the range 20~K $\ltsim$ T$_{d} \ltsim$
35~K.  Subsequent observations at mm wavelengths will be necessary to test this
assumption.    

Bothun $\&$ Rogers (1992) model the 100$\mu$m emission using an
isothermal dust slab at a dust temperature, T$_{d}$, that is optically thin to
far IR emission.  They assume that within each region of the disk sampled by
a pixel, radiative equilibrium exists between the interstellar
radiation field (ISRF) and the dust grains, and also that the ISRF,
represented by mean intensity, J$_{\nu}$, is uniform over the region.
They derive the following relationship between ${\it f}_{\nu_{0}}$, the observed flux
density in an IRAS passband with effective frequency $\nu_{0}$, and the
integrated mean intensity J:
\begin{equation}
{\it f}_{\nu_{0}}~\propto~{\cal F}_{\nu_{0}}(T_{d})\tau_{J} J,
\end{equation}
\noindent where $\tau_{J}$ is the mean-intensity-weighted optical
depth of the dust and
${\cal F}_{\nu_{0}}$ is the fraction of the total emission contained
within the IRAS band.  For a power law dust emissivity law that is proportional to
$\nu^{p}$, ${\cal F}_{\nu_{0}}$ can be expressed as
\begin{eqnarray}
{\cal F}_{\nu_{0}}~=~\int_{0}^{\infty} {{x^{3+p}}\over{e^{x}-1}}R_{x}dx/[\Gamma(4+p)\zeta(4+p)],
\end{eqnarray}
\noindent where $x=h\nu/kT_{d}$, R$_{x}$ is the system response
for the band that is tabulated in the IRAS Explanatory Supplement, and
$\Gamma$ and $\zeta$ are the Gamma and Riemann Zeta functions.

Because $\tau$ is proportional to mass in the optically thin regime, we can use the above
equations to trace dust mass as a function of radius if we know the dust
temperature profile, the 100$\mu$m light profile, and the integrated
mean intensity profile of the ISRF.  The radial profile of the dust mass
distribution can be derived without needing to derive the various coefficients which are difficult to
determine empirically.  We assume that
\begin{equation}
J~\propto~exp(-r/r_{s}),
\end{equation}
\noindent where $r$ is the radius and $r_{s}$ is the scale length of
the stellar component.  In the previous section, we determined that
the 100$\mu$m emission is also described by an exponential and thus
\begin{equation}
{\it f}_{\nu_{0}}~\propto~exp(-r/r_{d}),
\end{equation}
\noindent where $r_{d}$ is the 100$\mu$m emission scale
length.  
The only term in these equations that depends on the dust temperature
is ${\cal F}_{\nu_{0}}$.  Bothun $\&$ Rogers (1992) using 60$\mu$m and 100$\mu$m images for 28 nearby galaxies, found
that the dust temperature declines radially, due
to the exponentially declining ISRF.  These authors modeled the
temperature profile with an
exponential
\begin{equation}
T_{d}~\propto~exp(-r/r_{d}),
\end{equation}
\noindent which, for a limited range of scalelengths, can be approximated by a drop of 3 K for every optical
scale length (Bothun \& Rogers 1992).  This temperature profile is consistent with their
data out to a radius of $\sim$ 3 optical scalelengths, where the
temperatures converge to a near constant value of $\sim$ 28 K (Bothun
$\&$ Rogers).  
Therefore, we use the exponential temperature gradient, normalized to produce $T_{d}$ = 34.9 K at one optical
scale length and converging to a constant temperature of 28 K for $r >
3r_{s}$. 

We can now solve for the dust distribution once we adopt values for the
100$\mu$m emission scale length, $r_{d}$, and the scalelength of the
stellar component, $r_{s}$.  For $r_{d}$, we choose the mean of the
two best fit values, $r_{d}$ = 2.5~kpc.  As discussed previously, the
100$\mu$m scalelength could be equal or up to a factor of $\sim$2 smaller
than the stellar scalelength.  Thus, we investigate the regime $0.5 \leq
r_{d}/r_{s} \leq 1$.  When $r_{d}/r_{s}~=~1$, the model
predicts a slight fall in the dust mass distribution in the inner
regions of the galaxy (due to the radial temperature gradient)
and then the dust mass distribution flattens out when the temperature
converges to a constant.  In the absence of a temperature gradient, the distribution is flat at
all radii.  This result occurs because Eq. 1 simplifies to
$\tau \propto 1/{\cal F}_{\nu_{0}}(T_{d})$, for equal stellar and dust
emission scalelengths.  If
the temperature is constant with radius, then ${\cal F}_{\nu_{0}}$ is
constant, and $\tau$ is independent of radius.  For $0.5 \leq
r_{d}/r_{s} < 1$, an exponentially declining dust mass with radius is predicted, with
the dust being less centrally concentrated than the stars.  As
$r_{d}/r_{s}$ decreases, the dust mass scalelength decreases until it equals the stellar emission scalelength for
$r_{d}/r_{s} = 0.5$.

A flat or slowly declining exponential radial dust distribution is consistent with
the findings from optical studies.  de Jong (1996b) studied the optical and infrared
color profiles of 86 face-on disk-dominated galaxies using models of
well mixed stars and dust that include both scattering and
absorption.  He found that high central
optical depths and long dust scalelengths best reproduced the
observed color gradients.  However, he cautions that a realistic dust
distribution alone cannot
adequately account for the observed gradients, and that
stellar population characteristics, namely age and metallicity gradients, need
to be incorporated.  Peletier \etal (1995) also studied
spiral galaxies in the optical and infrared using models of smoothly
distributed dust and stars that only included absorption, but also
concluded that $r_{d}/r_{s} > 1$.  However, when they included scattering and
clumpiness in the dust models, $r_{d}/r_{s} \geq 1$ yielded acceptable
solutions.

\subsection{Implications of the Dust Mass Distribution}

How does the dust distribution compare to other galaxy components?
Neutral hydrogen has a radially flat distribution (Shaya $\&$ Federman 1987), while molecular gas appears to
decline exponentially (e.g.  Young $\&$ Scoville
1991).  Because our data cannot distinguish between a flat or
exponentially declining distribution of dust, we will discuss both
cases.  Some authors find a good correlation between
FIR luminosities and the distribution of molecular gas (e.g. Devereux $\&$
Young 1990a, 1990b, 1992, $\&$ 1993, Tinney et
al. 1990, Sanders \etal 1991, Downes \etal 1993, Sage 1993).  Devereux $\&$ Young
(1990a, 1990b, 1992, $\&$ 1993) studied the origin of the FIR
luminosity in IR luminous spiral galaxies and concluded that FIR luminosity is an
excellent indicator of star formation, being well correlated with
H$\alpha$ luminosity and the surface density of H$_{2}$ (as
inferred from CO observations) and poorly correlated with the surface
density of HI.  These authors claim that their observed FIR
luminosities are due to the reradiation by warm dust of
UV flux from young OB stars in HII regions, and therefore, the dust
exponentially declines with radius as do the stellar and molecular
components.   

Others provide
evidence for a second cooler, more diffuse cirrus-type component,
that is associated with the HI rather than star
forming regions and excited by the ISRF from old disk stars.  Using a model that includes both components, Lonsdale-Persson $\&$
Helou (1987) have calculated that cirrus-type emission can account for
up to between 50\% and 70\% of the total FIR emission from a median spiral
galaxy, and subsequent observations have supported their claim
(Walterbos \& Schwering 1987, Rice \etal 1990, Sauvage \& Thuan 1992,
Xu \& Helou 1996).  Furthermore, Sauvage \& Thuan (1992) observationally determined that
the fractional contribution of cirrus-type emission to the total FIR
of spiral galaxies is a function of type:  $\sim 3\%$ for Sdm,
increasing to $\sim 70\%$ for Sc galaxies, and reaching as high as
$\sim 86\%$ for Sa galaxies.  When normal spiral galaxies (i.e. not ultra luminous
IRAS galaxies nor active galaxies) are studied, the contribution from
the cirrus component becomes important and non-negligible
correlations between cold dust and HI are found, although cold dust is
better correlated with the total gas content (HI + H$_{2}$) than with either
phase separately (e.g.  Knapp, Helou, \& Stark 1987, Andreani, Casoli,
\& Gerin 1995, Xu \& Helou 1996).  Xu \& Helou (1996) studied the
diffuse component of the FIR emission of M31 by removing FIR emission
associated with HII regions and found a close association of the
diffuse cirrus-type component with HI, while a poor correlation was
found with H$_{2}$.  Interestingly, these authors calculate the radial
behavior of the V-band optical depth, $\tau_{V}$, and find that the
overall distribution of $\tau_{V}$ is rather flat out to a radius of
$\sim$ 14~kpc.  

Because the galaxies in our samples are
predominantly of type Sc and earlier, we expect the cirrus component,
rather than the dust component associated with HII regions,
to be the main contributor to the FIR emission.  However, a number
of authors have demonstrated that IRAS is not very sensitive to
emission by cold dust and that the emission from large amounts of cold
dust can easily be dominated by emission from a small amount of warm
dust (e.g. Bothun $\&$ Rogers 1992, Sage 1993, Block \etal 1994).
In reality, the 100$\mu$m emitting dust detected in our sample is probably associated, at some level, with both the molecular and atomic gas phases.  While our assumed dust
temperature of $\sim$30K is similar to dust temperatures derived for
samples of nearby spirals (e.g.  Young et al 1986, Walterbos \& Schwering
1987, Jura 1986, Bothun \& Rogers 1992), this is not the
characteristic temperature of either HII regions, which are hotter, nor the HI
interstellar medium, which is typically colder.  Our assumed dust temperature was
derived by Bothun \& Rogers (1992) using a single temperature model,
based on 60$\mu$m and 100$\mu$m data with very coarse resolution where
each pixel samples a wide range of local heating conditions, and
consequently is
probably an average of the temperatures of the two contributing components.  Therefore, although
our data do not allow us to differentiate between a flat and exponentially
declining distribution of dust, the 100$\mu$m emission measured in our
sample almost assuredly originates from both warm dust residing in HII
regions and associated with the exponentially declining molecular gas and also from the cooler,
cirrus component associated with the flatly distributed neutral hydrogen .

Ultimately, we would like to know
the radial extent of the dust layer.  In particular, does it have a
cutoff at some radius?  Unfortunately, our current data do not allow us to answer this
question, but we can use them to place a lower limit on that cutoff radius.  We have detected 100$\mu$m
emission from dust out to a radius of 27 kpc (with $\geq$
2$\sigma$ confidence).  This is a lower
limit to the cutoff radius and large amounts of cold dust
at larger radii could exist.  We can compare this limit to the known
edges of the stellar (e.g.  van der Kruit $\&$ Searle 1981a, 1981b,
1982a, 1982b) and neutral hydrogen (e.g.  Corbelli, Schneider \&
Salpeter 1989, Corbelli \& Salpeter 1993, Bland-Hawthorn, Freeman \&
Quinn 1997) disks.  van der Kruit \etal (1982a) found
that all of the stellar disks in their sample of 8 galaxies have
a sharp edge at 4~-~5 radial
scalelengths, or 14~-~25~kpc with an average value of 17~kpc.  Neutral
hydrogen disks also have a sharp boundary, although at radii beyond the edge of the stellar disk.
For example, galaxies in the Sculptor group have observed HI disks that extend
to $\sim$1.2R$_{25} \sim 23$~kpc (Bland-Hawthorn, Freeman $\&$ Quinn
1997).  Few galaxies have HI disks that extend farther to $\sim$2R$_{25}
\sim$~38~kpc (Bland-Hawthorn, Veilleux \& Carignan 1997,
Bland-Hawthorn, Freeman \& Quinn 1997).  The radial edge of our 100$\mu$m
detection, at a radius of 27~kpc, is beyond 4~-~5 stellar scalelengths
and therefore, beyond the edge of the stellar disk.  We do not
observe far-IR emission beyond the nominal edge of HI disks,
although because our detection is a lower limit, dust may extend
beyond the edge of the HI disk.

If dust is produced by stars, how can dust be found outside the
stellar disk?  One can speculate that it is primordial, i.e.  that it formed early in the process of galaxy
formation and then settled into a disk similarly to the HI, that galactic fountains, plumes of gas and dust ejected by supernovae into
the interstellar medium, pollute the outer galaxy, or that some disk
stars exist at these large radii.  Models of galactic fountains (e.g.  Corbelli
$\&$ Salpeter 1988) show that some of the hot material ejected by violent
supernovae, can rise many kiloparsecs into the halo and then
radiatively cool and fall back toward the outer part of the galactic
plane at distances $\geq$20~kpc.  A similar, but less energetic
mechanism of ejecting dust into the ISM involves radiation pressure
produced by
clusters of young OB stars.  Models of these dusty chimneys by
Ferrara (1995) show that some dust particles can attain rather high
velocities ($\sim$100~km/s) and reach distances of $\sim$3~kpc above
the disk plane.
However, the majority of the dust particles' motion is in a direction
perpendicular, rather than parallel, to the plane of the disk.
The redistribution of dust particles'
in Ferrara's models predicts a flattened dust distribution, which is
consistent with our work (see Fig~\ref{f7}).  Lastly,
although stellar disks are observed to truncate sharply, there is observational evidence for massive star formation in the outer regions
of galactic disks.  Hodge \etal (1988) and Davidge (1993) found blue
stars in two different fields in the outer disk ($\sim$20~kpc) of
M31.  If these
findings are representative of spiral galaxies, then there is at least
some star formation at the very edge of stellar disks which
could contribute dust at large radii.  Therefore, there are several
avenues that need to be more fully explored to determine the mechanism
involved in getting dust to r$\sim$30~kpc.

\section{Summary}

Using 100$\mu$m ISSA images and data from the RC3, we coadd the
flux from 299 galaxies and investigate the radial and azimuthal distribution of
100$\mu$m emission from dust in ensembles of galaxies.  The galaxies are divided into three
samples (where N~$=$ number of galaxies), according to their optical isophotal diameter, D$_{25}$:
(1)~~2.5$^{\prime} \leq$ D$_{25} \leq
5.0^{\prime}$~(N$_{2.5^{\prime}-5^{\prime}} = 184),~(2)~~5.0^{\prime}
\leq$ D$_{25} \leq 10.0^{\prime}~($N$_{5^{\prime}-10^{\prime}} = 91),$ and $(3)~~10.0^{\prime} \leq$ D$_{25} \leq
30.0^{\prime}~($N$_{10^{\prime}-30^{\prime}} = 24)$.  We find the following:

\noindent (1)  that 100$\mu$m emission extends out to
radii of 27~kpc at the 2$\sigma$ or greater confidence level for the
$5^{\prime} - 10^{\prime}$ sample and to
21~kpc for the $10^{\prime} - 30^{\prime}$ sample (no extended
emission is seen beyond the PSF for the $2.5^{\prime} - 5^{\prime}$ sample),

\noindent (2)  that the 100$\mu$m emission from both the $5^{\prime} - 10^{\prime}$ and $10^{\prime}
- 30^{\prime}$ sample is preferentially elongated along the major axis of the
disk, thus confirming the validity of the detections and demonstrating
that 100$\mu$m emitting dust is flattened into the disk,

\noindent (3)  that the 100~$\mu$m surface brightness
profiles from both samples can be well fit with an exponential with a dust
emission scale length of $\sim 2.5 \pm 0.8$~kpc, but there is evidence
for a second, more extended emission component, and

\noindent (4)  that a
model of the 100$\mu$m emission from an isothermal dust slab at
temperature, T$_{d}$, that is optically thin to emission can be used
to model our data (cf. Bothun \& Rogers 1992).  Because our
data do not allow us to calculate dust temperatures, we adopt the dust
temperature gradient derived by Bothun
\& Rogers (1992),
which can be approximated by a drop of 3~K for every optical scale
length, r$_{s}$, and is normalized by T$_{d} = $34.9~K at one optical scale length and
converges to a near constant value of $\sim$28~K for r$~>~3$r$_{s}$.
Using this model, we examine the range of acceptable dust mass
distributions allowed by our data and find that the dust is less
concentrated than the stars, with either a flat or slowly declining
exponential distribution. 

We have presented evidence for dust well beyond the stellar
disk, but have been unable to conclude whether dust exists
beyond the extent of the HI distribution.  Such dust, if present, should be quite
cold ($\ltsim$ 10~K) and difficult to detect with IRAS observations,
despite the possible evidence of a second component in our data.  The
future of this field relies on longer wavelength observations.
Preliminary detections of dust at large radii using 200$\mu$m
observations have recently been presented by Alton \etal (1998).  We
await further results from such observations.
 
\vskip 0.5in
\noindent
Acknowledgments:  DZ acknowledges funding from a NASA LTSA grant
(NAG-5-3501), a NSF
grant (AST-9619596), and the Packard Foundation.  AEN \& DZ gratefully acknowledge financial
support from the Cal Space Foundation.  AEN gratefully acknowledges
financial support from the University of California Graduate Research
Mentorship Fellowship program.  RMC acknowledges the support of the
Jet Propulsion Laboratory, Caltech, which is operated under contract
with NASA.

\clearpage
\centerline{References}
\refbook{Alton, P.B., Davies, J.I., Bianchi, S. \& Irewhella, M. 1998
in ``Galactic Halo: A UC Santa Cruz Workshop'', D. Zaritsky (ed.), (San
Francisco: ASP), p.149}
\aa{Andreani, P., Casoli, F. \& Gerin, M. 1995}{1995}{43}
\mnras{Andreani, P., \& Franceschini, A. 1996}{283}{85}
\apj{Bland-Hawthorn, J., Freeman, K.C. \& Quinn, P.J. 1997}{490}{143}
\refbook{Bland-Hawthorn, J., Veilleux, S., \& Carignan, C. 1997, in prep.}
\aa{Block, D., Witt, A., Grosbol, P., Stockton, A., Moneti,
A. 1994}{288}{383}
\apj{Bothun, G.D., Lonsdale, C.J., \& Rice, W. 1989}{341}{129}
\aj{Bothun, G.D., \& Rogers, C. 1992}{103}{1484}
\mnras{Clements, D.L., Andreani, P. \& Chase, S.T. 1993}{261}{299}
\apj{Corbelli, E. \& Salpeter, E.E. 1988}{326}{551}
\apj{Corbelli, E. \& Salpeter, E.E. 1993}{419}{104}
\aj{Corbelli, E., Schneider, S.E. \& Salpeter, E.E. 1989}{97}{390}
\apj{Davidge, T.J. 1993}{409}{190}
\apjlett{Devereux, N.A. \& Young, J.S. 1990a}{350}{L25}
\apj{Devereux, N.A. \& Young, J.S. 1990b}{359}{42}
\apj{Devereux, N.A. \& Young, J.S. 1992}{103}{1536}
\apj{Devereux, N.A. \& Young, J.S. 1993}{106}{949}
\apjlett{Downes, D., Solomon, P., Radford, S. 1993}{414}{L13}
\mnras{Fall, S.M. \& Efstathiou, G. 1980}{193}{189}
\refbook{Ferrara, A. 1995 in ``The Physics of Galactic Halos'', 156
WE-Heraeus-Seminar, H. Lesch \etal (eds.), (Verlag:Berlin)}
\apj{Fich, M., Blitz, L., Stark, A.A. 1989}{342}{272}
\apjlett{Helou, G. 1986}{311}{L33}
\apj{Hodge, P., Lee, M.G., Mateo, M. 1988}{324}{172}
\aa{de Jong, R.S. 1996a}{313}{45}
\aa{de Jong, R.S. 1996b}{313}{377}
\aa{de Jong, R.S. 1996c}{313}{557}
\apj{Jura, M. 1986}{306}{483}
\apj{Kennicutt, R.C. 1989}{344}{685}
\apj{Knapp, G.R., Helou, G., \& Stark, A.A. 1987}{94}{54}
\aa{van der Kruit, P.C., \& Searle, L. 1981a}{95}{105}
\aa{van der Kruit, P.C., \& Searle, L. 1981b}{95}{116}
\aa{van der Kruit, P.C., \& Searle, L. 1982a}{110}{61}
\aa{van der Kruit, P.C., \& Searle, L. 1982b}{110}{79}
\aa{Lequeux, J., Allen, R.J., Guilloteau, S. 1993}{280L}{L23}
\apjlett{Lin, D.N.C., \& Pringle, J.E. 1987}{320}{L87}
\apj{Lonsdale-Persson, C.J. \& Helou, G. 1987}{314}{513}
\apjlett{Low, F.J., $\it{et~al.}$ 1984}{278}{L19}
\apj{Maloney, P. 1993}{414}{41}
\refbook{Odenwald, S., Newmark, J. \& Smoot, G. 1996 preprint (astro-ph/9610238)}
\aa{Peletier, R.F., Valentijn, E.A., Moorwood, A.F.M., Freudling, W.,
Knapen, J.H., \& Beckman, J.E. 1995}{300}{L1}
\apj{Rice, W. 1993}{105}{67}
\apj{Rice, W., Boulanger, F., Viallefond, F., Soifer, B.T. \&
Freedman, W.L. 1990}{358}{418}
\apjsup{Rice, W., Lonsdale, C.J., Soifer, B.T., Neugebauer, G., Kopan,
E.L. et al 1988}{68}{91}
\mnras{Rowan-Robinson, M., \& Crawford, J. 1989}{238}{523}
\aa{Sage, L.J. 1993}{272}{123}
\apj{Sanders, D.B., Scoville, N.Z., \& Soifer, B.T. 1991}{370}{158}
\apjlett{Sauvage, M. \& Thuan, T.X. 1992}{396}{L69}
\apj{Shaya, E., \& Federman, S. 1987}{319}{76}
\apj{Sodroski, T. $\it{et~al.}$ 1994}{428}{638}
\apj{Tinney, C.G., Scoville, N.Z., Sanders, D.B., \& Soifer, B.T. 1990}{362}{473}
\refbook{de Vaucouleurs, G., de Vaucouleurs, A., Corwin, H.G., Buta, R.J.,
Paturel, G., \& Fouque, P. 1991, Third Reference Catalog of Bright
Galaxies (Springer, Berlin) (RC3)}
\aa{Walterbos, R.A.M. \& Schwering, P.B.W. 1987}{180}{27}
\refbook{Wheelock \etal 1994, IRAS Sky Survey Atlas Explanatory
Supplement, JPL Publication 94-11.}
\aa{Wouterloot, J.G.A., Brand, J., Burton, W.B., Kwee,
K.K. 1990}{230}{21}
\apj{Xu, C. \& Helou, G. 1996}{456}{163}
\apj{Young, J.S., Schloerb, F.P., Kenney, J.P., \& Lord, S.D. 1986}{304}{443}
\refbook{Young, J.S., Scoville, N.Z. 1991, Ann. Rev., 29, 581} 
\aj{Zaritsky, D. 1994}{108}{1619}

\end{document}